# Autonomous Cyber Defense Introduces Risk: Can We Manage the Risk?

Alexandre K. Ligo, US Army Corps of Engineers

Alexander Kott, U.S. DEVCOM Army Research Laboratory

Igor Linkov, US Army Corps of Engineers

Abstract: We discuss the human role in the design and control of cyber defenses. We focus in on machine learning training, algorithmic feedback, and algorithmic constraints, with the aim of motivating a discussion on achieving trust in autonomous cyber defenses.

Managing risks is a recurring research topic in cybersecurity, especially in the context of recent cyberattacks and ransomware-based disruptions in operations across the U.S. infrastructure [1]. Unfortunately, a growing number of malicious activities have been supported by machine learning (ML) such as password cracking, generation of new malware and deceiving biometric systems [2]. ML-based attacks are increasingly frequent and sophisticated, resulting in higher risk of damage to target systems. Moreover, in the world of cyber operations time is often measured in milliseconds. From denial-of-service attacks to spreading of ransomware or other malware across an organization's network, it is possible that manually operated defenses are not able to respond in real time at the scale required, and when a breach is detected and remediated the damage is already made.

Autonomous cyber defenses therefore become essential to mitigate the risk of successful attacks and their damage, especially when the response time, effort and accuracy required in those defenses is impractical or impossible through defenses operated exclusively by humans. Autonomous agents have the potential to use ML with large amounts of data about known cyberattacks as input, in order to learn patterns and predict characteristics of future attacks [3]. Moreover, learning from past and present attacks enable defenses to adapt to new threats that share characteristics with previous attacks.

On the other hand, autonomous cyber defenses introduce risks of unintended harm. Actions arising from autonomous defense agents may have harmful consequences of functional, safety, security, ethical, or moral nature. Such consequences can actually be considered negative externalities, or costs imposed on parties who do not agree to incur on them or do not benefit from the agents' actions [4]. For example, an autonomous defense agent might act against an attack on an oil refinery, but the defense causes an explosion killing residents nearby. Or algorithmic biases might result in defenses that prioritize profits over safety or equity.

What can be done to increase benefits associated with autonomous defenses and reduce the associated risk of negative externalities? Here we discuss possible approaches to building autonomous cyber defenses that provide cybersecurity while minimizing risks. In particular, we discuss the human role in the design and control of defenses, which include ML training, algorithmic feedback and algorithmic constraints. While this discussion is non-exhaustive, it provides possible directions of research in risk management for autonomous defenses.

## Should humans interfere with the operation of autonomous defenses?

A natural way to address the risk of negative externalities is to have autonomous agents "team up" with human operators. We argue that such an approach may cause more harm than good under a number of important circumstances.

Consider the related but perhaps familiar context of autonomous driving. The Society of Automotive Engineers (SAE) defines a five-level scale of automation [5]. In all but level 5, a human driver is expected to take control over the machine during an emergency. Consider the scenario of a self-driving car in level 3 or 4, thus having a human driver in stand-by, when a child darts across the street from between parked cars to retrieve a ball. If the human tries to retake control to swerve and miss the child in its path the vehicle could override the driver. If the vehicle senses the child and begins a collision-avoidance maneuver, then any human operator action may ruin the automated system's plan for avoiding a collision, or the person's reaction time may be dangerously longer than the time taken by the machine. If neither the human nor the vehicle does anything, there will be a dead child and liability for all involved. As long as the probability of error by the vehicle is small, the best course of action is that the human driver does **not** interfere with the autonomous operation. In fact, this is the goal of some of the developers of self-driving technology [6].

The car example has similarities with autonomous cyber defenses teaming with humans. For example, both types of systems require quick and accurate decision making. If machine action (either assisted by human or not) is not effective, negative externalities may arise. However, there is a key difference. Autonomous cars are largely designed to replace a human ability – driving. Therefore, a trained human driver can usually take the wheel and achieve currently acceptable driving performance if enough time is available. On the other hand, autonomous cyber defenses may need to perform at "super-human" levels, because they are required to either achieve response times, process volumes of data, or respond at levels unachievable by human operators. Moreover, if a defense switches from purely autonomous to purely human, it is likely that the human defense will not promptly have the level of situation awareness [7] required to respond safely. These attributes of autonomous defenders make it even more risky for humans to intervene in their operation.

Therefore, we do not recommend human interference with the operation of autonomous cyber defenses. Nevertheless, there are still ways in which humans and machine can and should collaborate to minimize risk.

## Human intervention in training rather than operation

One alternative to human-in-the-loop in cyber defense *operation* is to enhance human intervention in the *training* of autonomous defenses, to mitigate risks of defense agents. Cybersecurity can be enhanced by design when humans actively shape an agent's behavior. Some forms of ML imply human intervention when AI algorithms of agents are trained *before* defenses are deployed. Humans can and should interfere with the learning of cyber defense agents, modulating the agents' actions through functional, safety, security, ethical, and moral considerations.

More specifically, this kind of human-in-the-loop intervention materializes in the *training* phase of ML. Goodfellow et al. argue that the most successful form of ML is supervised learning [8], in which a training dataset of known outcomes (or *labeled examples*, in ML jargon) is used to determine the parameters of the application. One common example of supervised learning is the email spam filter, where a list of emails previously classified as spam (by human beings) is used for training. The ML classifier then uses the train dataset to "learn" the typical characteristics of spam email, enabling the system to successfully classify new incoming email as legit or spam. This example illustrates the role of humans, which is analogous in the training of autonomous defenses through carefully labeling. "Supervised learning by definition relies on a human supervisor to provide an output example for each input example" [8]. Human supervision is key not only to provide representative examples in the train dataset, but perhaps more importantly, to provide examples that mitigate risk of harmful outcomes by the AI defender.

This idea is simple, but its implementation is challenging. For example, in the self-driving scenario the system would need ML algorithms that classify driving decisions as having high or low risk of causing negative externality (e.g., killing a pedestrian). Such models are probably less mature than ML algorithms that detect lanes or traffic signals. Moreover, it is unclear where the labeled examples of high/low risk would come from. Training autonomous cyber defenses is likely better known for intrusion detection than to predict the risk of negative externalities, and finding labeled data related to negative externalities may be difficult in the near future.

## Beyond supervised learning: human-in-the-loop through hierarchical feedback

One limitation of training autonomous defense agents through supervised learning is its dependency on known examples. Spam filters are only capable of detecting emails that share key characteristics with spam that were previously labeled as such. It is of little use to train agents against novel attacks or malware that can conceal its known "signatures." Likewise,

………………………….

important content that was not previously labelled as so may be classified as spam, which can be a negative externality. In this sense, the agents' potential to enhance cybersecurity while reducing risk of negative externality is limited by the data.

Hence, human intervention should go beyond training. We have argued that human intervention should *not* include the *operation* of AI defense agents. Rather, humans can team up with autonomous cyber defenses in a hierarchical arrangement. While the system would quickly produce purely autonomous responses to attacks, humans could evaluate such responses (probably not in real time) and adjust the system to produce responses in the future with lower risk of negative externality. This is consistent with the view of Kott and Alberts [9]: the superior response time and data processing of cyber systems might be combined with human judgment of safety or ethical consequences.

One possible direction of research is on reinforcement learning [10], such as proposed conceptually by Cam for cybersecurity with autonomous agents [11]. In general, reinforcement learning refers to ML algorithms that interact with the environment in a recurrent or "trial-and-error" fashion where the system takes an action, assesses the consequences of that action in the form of a "reward," and then changes its behavior to maximize the reward from future actions. Cam discusses how reinforcement learning can be used to predict actions from attackers and enable an AI defender to counteract accordingly. The proposal does not address the mitigation of risk of negative externalities caused by the agents. Nevertheless, this type of work lays a direction for future research on having the risk mitigation as a component of reward in the reinforcement loop.

Another ML approach that may be useful to mitigate externality risk is the use of Generative Adversarial Networks (GANs). This is a category of ML methods that has gained popularity in part because of its accuracy in certain applications, and in part because it may help overcome one limitation of supervised learning. "In order to reduce both the amount of human supervision and the number of examples required for learning, many researchers today study unsupervised learning, often using generative models" [8]. In general, GANs refers to ML based on the interaction of two "competing" systems: a generative algorithm and a discriminative algorithm. Labeled, training examples are input to the generative model, which determines its parameters to approximate the (unknown) distribution of examples as closely as possible. On the other hand, the goal of the discriminative algorithm is to classify if given examples came from the training dataset or the generative system. The two algorithms play a game where the generative system attempts to synthetize data that mimics the training dataset and the discriminative algorithm tries to detect the "fake" examples. This adversarial game proceeds until the discriminative system can no longer differentiate the output of the generative algorithm from the training dataset [8], [12], [13]. GANs have proved potential to generate false images and videos in deepfakes [8]. A known example of research on GANs applied to cybersecurity is on password cracking [2].

One potential direction for research can be the use of GANs in a way that the generative system creates possible defenses (based on a training set of optimal autonomous defenses) and the discriminative algorithm determines whether the generated defenses deviate from functional, safety, ethical, and moral criteria embedded in the training examples. In this case, a well-tuned GAN will elect the least risky defenses with respect to the rules. Like the supervised learning idea mentioned above, the GAN approach would depend on training data that represent a space of possible actions with autonomous agents and conditions, with labels indicating what actions maximize functional, safety, ethical, and moral goals under given conditions. These data are probably costly to obtain and label, but this is not unusual with AI. The GANs approach is a way to have the human-in-the-loop not directly in operations. This is because humans will fulfill the important role of labeling risk-mitigating examples of defenses.

## Algorithmic constraints

We discussed the use of supervised, reinforcement or adversarial learning to improve autonomous agents' design and mitigate risk of negative externalities arising from their actions. In any case, research is needed to assess the degree in which ML algorithms satisfy predefined constraints that prevent agents to act against functional, safety, security, ethics, and moral goals.

Another form of hierarchical team up (i.e., with humans collaborating but *not* operating) can be through constraints.  As an analogy with Asimov's three laws of robotics [14], constraints may be imposed at the design phase in such a way that if the behavior learned from data by the AI agent violates those rules, then the agent's decision is aborted or reversed. For example, a cyber defense agent might learn to shut down an oil pipeline in the event of unauthorized access. But what if that line is critical for heating to a certain population in the winter? The agent's action would jeopardize the system purpose.

While the idea of constraints may look simple, its implementation is challenging. Understanding (and perhaps mitigating) harmful consequences of AI is an interdisciplinary effort that includes areas such as machine ethics [15] and AI governance [16]. Moreover, some ML methods raise issues of model explainability. Methods such as deep neural networks produce results in which the patterns learned from data are not obvious or intuitive for human observers. As a consequence, trust in their results is often questioned. Research is needed to verify the compliance of defense agents' outcomes to functional, safety, security, ethics, and moral constraints. For example, trust in rule compliance can be built in a way similar to how trust is achieved by humans [17]. One way to trust an individual is to assess their history of transactions. Likewise, developers of self-driving cars have been accumulating millions of miles driven autonomously [18] to build trust among developers, regulators and the general public. In a similar fashion, we propose that the "action history" of an AI defense agent be gauged, first in a simulated or test environment, to estimate its compliance to rules whenever the agent acts and thereby build trust in not causing negative externalities.

……………………………….

## The way forward

Attacks by adversarial systems with AI capability have become increasingly frequent and sophisticated [19]. Autonomous defense agents are fundamental to enhance cybersecurity. However, such defenses may increase the risk of negative externalities, or the collateral harm to systems or individuals not directly involved in the cyber operations in question. To mitigate those risks, it is fundamental that AI defense agents and humans coordinate to incorporate functional, safety, security, ethics, and moral considerations into their design, models, and decisions. However, we argue that human intervention in the operation of autonomous cyber defenses may cause more harm than good. Possible alternatives include varying degrees of human intervention in the machine learning process, as well as the enforcement of algorithmic constraints to mitigate risks arising from the agents' actions. These approaches urge interdisciplinary collaboration to deploy defense agents that are accurate and trustworthy. Moreover, it is important to take into consideration the role of the organization's management and leadership to assess and mitigate the risks of negative externalities arising from autonomous cyber defenses. To this effect, managers have a pivotal role in going beyond a pure cost-minimization mindset and committing enough resources to plan for worst-case scenarios, to design defenses that avoid unintended harm, and to process large amounts of data (both from simulated and real operations) to adjust future behavior of the defenses.

While autonomous defenses should minimize new risks, their design may also incorporate other features such as resilience. Risk assessment and management methodologies typically focus on managing threats to avoid disruptions or hardening the system to decrease its vulnerability. A potentially more comprehensive approach is to focus on *cyber resilience*, or as defined by the National Academy of Sciences, a system's ability to: (i) plan and prepare responses to threats in order to keep assets available and systems operating, (ii) absorb threats in a way that minimize the impact on critical functionality and systems goals, (iii) recover system functionality to pre-disruption levels as quickly as possible, and (iv) adapt systems, personnel and processes to better respond to future threats [20], [21]. Therefore, the resilience approach goes beyond trying to make a system impenetrable. Rather, a resilient system absorbs disruptions gracefully and recovers quickly from them. Human-system team up could go beyond hardening the system while minimizing risk to effectively enhance other cyber-resilience abilities such as *response* and *adaptation* after an attack materializes. To this effect, research is needed on models that measure risk and cyber resilience, using measurements in a hierarchical or supervisory loop to change future actions (and further enhance resilience).

## Disclaimer

………………………….

*Vita*: Alexandre K. Ligo is a Research Scientist with the Engineer Research and Development Center (ERDC), US Army Corps of Engineers (USACE), Concord, MA 01742, and the University of Virginia, Charlottesville, VA 22904, USA. Contact him at aligo@virginia.edu.

*Vita*: Alexander Kott is the Chief Scientist of the U.S. DEVCOM Army Research Laboratory, Adelphi, MD, 20783, USA. Contact him at alexander.kott1.civ@mail.mil.

*Vita*: Igor Linkov is a Senior Science and Technology Manager with the ERDC-USACE, Concord, MA 01742, USA. Contact him at igor.linkov@usace.army.mil.